



\message{<< Assuming 8.5" x 11" paper >>}    

\magnification=\magstep1	          
\raggedbottom

\parskip=9pt

%

\def\singlespace{\baselineskip=12pt}      
\def\sesquispace{\baselineskip=16pt}      









\font\openface=msbm10 at10pt
 %

\def\Minkowski     {{\hbox{\openface M}}}

\def\sqr#1#2{\vcenter{
  \hrule height.#2pt 
  \hbox{\vrule width.#2pt height#1pt 
        \kern#1pt 
        \vrule width.#2pt}
  \hrule height.#2pt}}


\def\dal{\mathop{\,\sqr{7}{5}\,}}



\def\lto{\mathop
        {\hbox{${\lower3.8pt\hbox{$<$}}\atop{\raise0.2pt\hbox{$\sim$}}$}}}
\def\gto{\mathop
        {\hbox{${\lower3.8pt\hbox{$>$}}\atop{\raise0.2pt\hbox{$\sim$}}$}}}
%
%
%



\def\part{\subseteq}		



\def\to{\mathop\rightarrow}	


\def\SetOf#1#2{\left\{ #1  \,|\, #2 \right\} }



\def\interior #1 {  \buildrel\circ\over  #1}     




\def\basisvector#1#2#3{
 \lower6pt\hbox{
  ${\buildrel{\displaystyle #1}\over{\scriptscriptstyle(#2)}}$}^#3}









\fontdimen16\textfont2=2.5pt
\fontdimen17\textfont2=2.5pt
\fontdimen14\textfont2=4.5pt
\fontdimen13\textfont2=4.5pt




%



%
%
%
%
%
%
%
%
%
%
%
%
%
%
%

%
 \let\miguu=\footnote
 \def\footnote#1#2{{$\,$\parindent=9pt\baselineskip=13pt%
 \miguu{#1}{#2\vskip -7truept}}}
%
%

\def\linebreak{\hfil\break}
\def\lbr{\linebreak}
\def\pagebreak{\vfil\break}


\def\BulletItem #1 {\item{$\bullet$}{#1 }}
\def\bulletitem #1 {\BulletItem{#1}}

\def\author#1 {\medskip\centerline{\it #1}\bigskip}

\def\address#1{\centerline{\it #1}\smallskip}

\def\furtheraddress#1{\centerline{\it and}\smallskip\centerline{\it #1}\smallskip}

\def\email#1{\smallskip\centerline{\it address for email: #1}} 

\def\AbstractBegins
{
 \singlespace                                        
 \bigskip\leftskip=1.5truecm\rightskip=1.5truecm     
 \centerline{\bf Abstract}
 \smallskip
 \noindent	
 } 
\def\AbstractEnds
{
 \bigskip\leftskip=0truecm\rightskip=0truecm       
 }

\def\section #1 {\bigskip\noindent{\headingfont #1 }\par\nobreak\noindent}

\def\subsection #1 {\medskip\noindent{\subheadfont #1 }\par\nobreak\noindent}
 %

\def\ReferencesBegin
{
 \singlespace					   
 \vskip 0.5truein
 \centerline           {\bf References}
 \par\nobreak
 \medskip
 \noindent
 \parindent=2pt
 \parskip=6pt			
 }
 %


\def\reference{\hangindent=1pc\hangafter=1} 

\def\ref{\reference}

\def\sepref{\parskip=4pt \par \hangindent=1pc\hangafter=0}
 %

\def\journaldata#1#2#3#4{{\it #1\/}\phantom{--}{\bf #2$\,$:} $\!$#3 (#4)}
 %

\def\eprint#1{{\tt #1}}

\def\arxiv#1{\hbox{\tt http://arXiv.org/abs/#1}}
 %

\def\webtilde{\lower2pt\hbox{${\widetilde{\phantom{m}}}$}}

\def\webhome{{\tt {http://www.physics.syr.edu/}{\webtilde}{sorkin/}}}
 %

 %


\def\hpf#1{\webhome{\tt{some.papers/}}}
 %

\def\hpfll#1{\webhome{\tt{lisp.library/}}}
 %




\font\titlefont=cmb10 scaled\magstep2 

\font\headingfont=cmb10 at 12pt
%

\font\subheadfont=cmssi10 scaled\magstep1 
%








\overfullrule=0pt



\phantom{}




\sesquispace
\centerline{{\titlefont  Light, Links and Causal Sets}\footnote{$^{^{\displaystyle\star}}$}%
{ Published in the proceedings of the DICE2008 meeting, held Sept. 21-27,
  2008 in Castiglioncello, Italia: 
 \journaldata{Journal of Physics}{Conf. Ser. 174}{012018}{2009}
}}

\bigskip


\singlespace			        

\author{Rafael D. Sorkin}
\address
 {Perimeter Institute, 31 Caroline Street North, Waterloo ON, N2L 2Y5 Canada}
\furtheraddress
 {Department of Physics, Syracuse University, Syracuse, NY 13244-1130, U.S.A.}
\email{sorkin@physics.syr.edu}

\AbstractBegins                              
After sketching a context in which to seek observable signals of
spatio-temporal discreteness, I briefly review the status of the causal
set program for quantum gravity, concluding with a simple model for the
field produced by a moving charge in a background causal set.
\AbstractEnds



\sesquispace


\section{1.~ How might discreteness show up?}
Riemann had a nice phrase somewhere that talked about ``the reality that
underlies spacetime.''\footnote{$^\star$}
{Of course he would have said ``space'' rather than ``spacetime''}
%
%
What we know of quantum gravity
suggests that that reality is more in the nature of a discrete structure
than a continuous one.  If so then it's important to think about
phenomena that could reveal this underlying discreteness.
Several such possibilities come to mind, some of them present by
definition, others suggested by analogy with the discreteness of
ordinary material, and still others specific to one or another of the
deep structures of spacetime that have been proposed by workers on
quantum gravity. 

Almost by definition, discreteness implies a ``cutoff'' in energy and
wavelength.  However (unless the idea of ``large extra dimensions''
turns out to be correct, cf. [1]) the expected
Planckian scale of this effect places it beyond {\it direct} observation
for the present.  Moreover (and this is why I included the qualification
``Almost'' just above) the concept of a cutoff takes on a more subtle
meaning in the context of a discreteness that respects Lorentz
invariance, like that of the causal set.  One cannot naively translate
such a granularity into the statement that an upper limit to frequency
would exist in any given frame of reference.  Such a claim would
conflict with the fact, that, for example, a plane-wave solution of the
massless wave equation can have any frequency at all, depending on which
frame it is referred to.\footnote{$^\dagger$}
{Here I am of course identifying the physical symmetry group of  Minkowski
 spacetime with the Poincar{\'e} transformations as normally defined.
 This ignores the possible extension of the concept of symmetry to a
 Hopf algebra that is not a group algebra, as essayed in ``deformed
 special relativity'' and non-commutative geometry more generally.
 [2]}
This subtlety does not mean that a Lorentz-respecting cutoff must remain forever
elusive, of course, only that it can be detected solely by effects whose
nature is intrinsically frame-independent.  For example, the notion of
center of mass energy is invariant, and correspondingly, one can easily
see in the context of (background) causal sets that scattering cross
sections must fall to zero at Planckian energies.  One could in principle
accelerate particles to arbitrarily high energy (limited only by
infrared cutoffs of cosmological origin), but if such ``transplanckian''
particles were to collide, the associated wave packets would simply
``pass through'' each other.  They would not ``see each other'' because
they would be supported on disjoint sets of causet elements.

Interesting as this 
``asymptotic freedom'' 
effect is in principle, it seems far removed from
observability in practice, and people have sought elsewhere for the
phenomenology of spatio-temporal discreteness.  By way of inspiration,
one can call to mind some of the analogous phenomena that testify to the
atomicity of ordinary matter.  Within this ambit, the effect that has
received the most attention is perhaps the possible breaking of Lorentz
invariance.
Here the condensed-matter analogs are the partial breaking of
translational and rotational invariance by a crystal (resulting for
example in Umklapp scattering processes and the M{\"o}ssbauer effect)
and the related high frequency modification of dispersion relations for
waves traveling through the crystal.  However, the observational and
experimental evidence to date speaks against any such failure of Lorentz
invariance, suggesting that we should limit our search to phenomenology
consistent with this symmetry.  
Indeed, the vista from causal sets suggests that this negative verdict
will persist indefinitely, because it seems to be impossible to design a
causet that would single out any frame of reference without coming into
contradiction with the basic hypothesis that ``number = volume''.  (More
specifically, it seems that the most uniform distribution of points in 
Minkowski spacetime $\Minkowski^4$ is given by a Poisson distribution;
and one knows that Poisson distributions are frame-independent in a 
strong sense [3].)

But, not all matter is crystalline, and dispersion relations are not the
only --- or even the best --- messengers from the realm of atoms.  In
gases and liquids, another whole family of phenomena springs from the
irregular arrangement and chaotic motions of the constituent molecules.
Some familiar instances of these ``fluctuation phenomena'' 
are the Brownian motion of microscopic particles suspended in a medium
and the scattering and extinction of light propagating through air.  By
way of
analogy people have asked whether particles moving freely through the
vacuum might deviate from geodesic motion, and whether light from
cosmologically distant sources might be attenuated or deviated or
otherwise affected by a Planck-scale discreteness [4].

To fully answer these questions is not possible outside of a completed
theory of ``quantum gravity'', but that need not stop us from
seeking partial answers based on plausible models of an incomplete
and/or partly phenomenological nature.  Below I will briefly present one
such model [5]
that describes the propagation of light (more precisely, a
massless scalar field) through a causal set as a process of ``direct
transmission from source to sink.''  Limited though it is, this model
lets us appreciate that neither modified dispersion relations nor loss of
coherence is a necessary consequence of 
discreteness.  It
will also allow one to estimate the size of some of the fluctuations
in signal strength that do result from the discreteness.
Before getting to that, however, I'd like to recall two or three more
types of phenomena that have been suggested as potential ``signals'' of
a quantum gravity founded on an atomistic deep structure of spacetime.

The first of these is the causal set counterpart of Brownian motion, but
a similar effect could be expected in any Lorentz-symmetric discrete
theory.  Precisely because of the absence of a distinguished frame,
a direct analog of the Wiener process is ruled out
in the sense that fluctuations in the position cannot not be
Markovian;
but fluctuations of velocity or momentum pose no such
difficulty.  In fact, one can define a spacetime analog of the
``Ornstein-Uhlenbeck'' version of Brownian motion, in which the
particle's worldline is differentiable but its tangent-vector is not,
the acceleration being subject to a sort of Lorentz-invariant white
noise.  For a massive particle the resulting stochastic process
[6]
[7] [8] is
characterized by a single phenomenological parameter, a diffusion
constant with the dimensions of inverse proper time.\footnote{$^\flat$}
{These are the dimensions if the diffusion is referred to
 velocity-space.  If it is referred to momentum-space the dimensions are
 $mass^2/time$.} 
For a
massless particle, a second free parameter enters (a drift coefficient in
energy), while for a photon (massless with polarization), still other
parameters become relevant [8].  Since we have no
estimate of any of these phenomenological  parameters from first principles,
the best that one can do is to seek observational bounds on them.  Such
bounds have been deduced from the undistorted Planckian shape of the
cosmic microwave background (CMB) spectrum [8] and from
cosmological bounds on the kinetic energy carried by relic neutrinos
[9]. 

The second type of phenomenon stemming from the discreteness of the
causet is a potential failure of locality at sufficiently high energies.
At the root of this possibility is an inner contradiction among
discreteness, Lorentz invariance, and locality: any two can coexist but
not all three together.\footnote{$^\star$}
{To get a feel for this contradiction, observe that two
 ``infinitesimally nearby'' point-events are ``infinitesimally nearby''
 in all frames, but two points at a small but finite timelike distance
 in a given frame will be separated by arbitrarily great times in other
 frames.}
Some models of discrete spatio-temporal structure maintain an
(approximate) locality at the expense of Lorentz invariance.  But with
causal sets, this option is difficult or impossible to implement (even
if it were desirable), and the problem becomes how to recover at least
the appearance of locality at sufficiently great length scales, where a
continuum spacetime offers a good approximation to the underlying
causet.  Work on this problem has illustrated concretely how a
coarse-grained locality on mesoscopic scales can emerge from microscopic
interactions that are radically nonlocal, but it also suggests that the
nonlocality might persist at length scales much greater than the
fundamental discreteness scale (Planck scale) [10].  Were evidence
of such nonlocality discovered, that would open up an entirely new
phenomenological area for quantum gravity.  However I don't know that
anyone has proposed a clear experimental signal of nonlocality that one
could easily look for.


A final phenomenon that deserves incidental mention here --- concerning
the cosmological constant $\Lambda$ --- differs from the others I've
been referring to in that it pertains to full quantum gravity, as
opposed to the more limited setting of the propagation of fields and
particles on a fixed, non-dynamical background.  In conjunction with the
quantum mechanical conjugacy between $\Lambda$ and spacetime volume $V$,
Poisson fluctuations in the ``density of causet elements'' imply
fluctuations in $\Lambda$ with an order of magnitude $1/\sqrt{V}$.  This
heuristic argument [11] led to the prediction, which later
came true, that the contemporary value of $\Lambda$ would be found to be
comparable in magnitude (but not necessarily in sign) to the density of
baryonic and other ``matter''.  A simple model built on the above idea
can be found in [12], but a fully consistent
phenomenological theory of such fluctuations remains to be devised.

\section{2.~ The causal set program}         
In the preceding introduction, I have sketchily reviewed some possible
phenomenology corresponding to the hypothesis that a discrete structure
supersedes the Lorentzian manifold at Planckian scales.  Partly to
provide background for those remarks (and for the remarks which follow),
and partly in obedience to a request from the conference organizer, I
will try in the present section to review --- even more sketchily ---
the overall status of the causal set program itself\footnote{$^\dagger$}
{For some general references to this program see  [13].}

At the root of this program is the recognition that, when it is combined
with volume information, causal structure alone suffices to reproduce
fully the geometry of spacetime.  In a continuum --- what Riemann
called a ``continuous manifold'' --- volume information is lacking.  But
what Riemann called a ``discrete manifold'' does carry an obvious
volume-measure obtained simply by counting elements.  In this case
``less is more, as well as less''.  The kinematic meaning of the causet
hypothesis, then, is that macroscopic spacetime $M$ is an effective
description of (an approximation to) a causal set $C$.  Mathematically
$C$ is a partial order or ``poset''.  Its defining order relation
$\prec$ gives rise to the light-cone structure of spacetime, while its
innate volume measure gives rise to the continuum volume element
$\sqrt{-g}d^4x$.  In brief ``Number = Volume'', and thus, ``Order +
Number = Geometry''.

A more precise notion of the causet-spacetime correspondence has been
based on the concept of ``faithful embedding''.\footnote{$^\flat$}
{With Lorentzian geometry, the concept of ``closeness'' between two
 spacetimes $M$ or between a spacetime and causet $C$ turns out to be
 much more delicate than one would have been led to believe by
 the example of Euclidean signature.}
One says that $f:C\to{M}$ is faithful if the point-set $f[C]$ ``could
have been obtained by Poisson-sprinkling points into $M$ at unit
density''.  It it this notion of correspondence that is employed in the
phenomenological applications discussed in the previous and following
sections. 


What would it take to turn the causet hypothesis into a theory of quantum gravity?
For a physicist, it seems natural to follow the scheme advocated by
Taketani that views a theory as comprised of three components or
``stages'', which one may call kinematical, dynamical and
phenomenological.  For causets, kinematics refers first of all to the
{\it kind} of structure one has, i.e. to the definition of a causet as a
locally finite partial order, and more broadly to the development of the
mathematical ideas proper to causets (the mathematics that would play
the same role for causet theory that differential geometry plays for
general relativity).  Of special importance is the development of a
``dictionary'' allowing one to translate back and forth between order
theoretic and geometric concepts.  By dynamics I mean what one might
describe as the ``equations of motion'' of the causet.  But in addition
to this (which pertains to full quantum gravity) there are also
questions analogous to those belonging to the theory of quantum fields
on a background spacetime.  With the exception of the above remarks on
$\Lambda$, all the various effects mooted above fit into this second
category.  The word phenomenology, finally, needs no definition and no
separate discussion in this section, since the other two sections of the
paper are devoted to it.\footnote{$^\star$}
{A ``phenomenological'' application of causets not mentioned in
 those sections concerns black hole entropy.  Unfortunately
 the available accounts are rather outdated by now [14].}

\subsection {Kinematics}

At the level of kinematics, much is known, although some important
questions still remain open.  Some of the necessary concepts can be
borrowed from the mathematical discipline of combinatorics (for example
the ``height of an interval''), others can be copied from the ``global
causal analysis'' of continuum general relativity, while still others
(like Myrheim's ``ordering fraction'', or ``coarse-graining'' by random
selection of a suborder) seem not to have been studied before.  As
things stand now, one has a growing dictionary allowing one to ``read
out'' geometrical and topological information from order
information.\footnote{$^\dagger$}
{Of course this type of translation only makes sense when the causet in
 question is in fact well approximated by some Lorentzian manifold.}
Corresponding to the timelike distance between two points of spacetime
is the length of the longest (not shortest) chain between two causet
elements.\footnote{$^\flat$}
{This estimator turns out to be remarkably accurate.  There is some
 evidence that its fluctuations grow only {\it logarithmically} with the
 number of elements, in dimensions $3+1$ and higher!}
(As its name suggests, a {\it chain} in a partial order is a sequence of causet
elements, $e_1\prec e_2\prec e_3,\cdots \prec e_n$,
each an ``ancestor'' of the next one in the sequence.)
Corresponding to spacetime dimensionality we have a number of
different estimators.  Here I will just mention three.  For a given
interval $I$ we have first of all the Myrheim-Meyer dimension, which
seems to be the most accurate so far [15].
It is a certain function of the
ordering fraction $R/{N \choose 2}$, $R$ being the number of related
pairs of elements of $I$ and $N=|I|$ being simply the number of elements
of $I$.  (An {\it interval} in a poset is the set of elements causally
between two given elements $a$ and $b$: $I=\SetOf{x\in C}{a\prec x\prec
b}$.)  A second estimator, the ``midpoint scaling dimension'' of an
interval is analogous to a Hausdorff or fractal dimension.  A third
estimator suggested by Eitan Bachmat [16], 
can be applied directly to the
causet as a whole, not just to intervals within it.  It also is a kind
of scaling dimension, based in this case on how the ``height'' of $C$
(the length of the longest chain it contains) scales under
coarse-graining.  All these estimators (for timelike distance and
dimension) have been tested in a range of examples, but rigorous proofs
of their validity exist only in special cases, like that of Minkowski
spacetime. 


In principle, if one knows all the distance relations in a Lorentzian
manifold $C$, one know everything there is to know about $M$, including
its topology.  Thus, one can expect that something similar would be true
of a causet well approximated by $M$: one could deduce the topology
approximately by counting longest chains within $C$.  However, no
technique for doing this is known.  Nonetheless, there has been recent
progress in recovering algebraic-topological information directly, more
specifically in assigning homology groups to ``slices'' (the discrete
analogs of spacelike hypersurfaces)
[17] [18].
There also exist proposed definitions for homology groups pertaining to
$C$ as a whole (as opposed to slices within it), but they remain
untested.

Probably the most exciting of the recent kinematical developments, is
the prospect of an estimator for the Ricci scalar $R$.  Were this
prospect to pan out, it would furnish a causal set counterpart to the
Lagrangian of continuum gravity, and therefore a direct route to setting up a
``path integral'' for causet dynamics.  The estimator in question grew
out of the attempt, referred to earlier, to formulate an approximately
local analog of the scalar wave equation for a field $\phi$ defined on a
causet [10].  The resulting ``discrete D'Alembertian operator''
$\dal$ does yield such an analog, and it can indeed be used to study
propagation of waves in a causet.  But once one has it, $\dal$ can also
be combined with the ``Synge world function'' $\sigma$ to obtain a
discrete analog of $R$, by forming (purely at a kinematical level) the
expression $\dal\dal\sigma$.  In the continuum, this combination is
known to reduce to (a multiple of) $R$ in the coincidence limit.  The
question now is whether it will do likewise in the discrete case.

A final kinematical result that bears mention here relates back to the
question of Lorentz breaking and the causet-spacetime correspondence.  I
asserted earlier that discreteness of the causet type respects the full
symmetry group of $\Minkowski^4$, including the Lorentz boosts.  This
claim rests on the fact that the Poisson ``sprinkling'' process is
Lorentz invariant (which invariance rests in turn on the fact that
4-volume is invariant).  What this really signifies is that, given a
large number of sprinklings into $\Minkowski^4$, one could not deduce a
distinguished reference frame from the entire collection.  Logically
this still leaves open the possibility that
even though the {\it process} of sprinkling is invariant 
an {\it individual} sprinkling
would still determine a frame, so that invariance could return only in some
sort of ``sum over causets''.  Were things to turn out this way, the
project of studying wave or particle propagation in a {\it single}
causet, taken as typical, would be misguided.  In effect, one would have
to go all the way to full quantum gravity in order to analyze the
effects of discreteness on propagation.  Fortunately, one can prove that
(with probability 1) even a single sprinkling fails to determine a frame
[3].  Accordingly, it makes sense, as a first effort,
to study propagation in a single sprinkled causet, and this is what we
will do in Section 3.

\subsection {Dynamics}
In relation to causal sets, the word dynamics refers most fundamentally
to the laws or principles that would govern the structure of the causal
set itself, the quantum causet replacement for the Einstein equation.
%
%
But dynamics can also be taken in a more limited sense, to refer to any
process of evolution of fields, particles, or other forms of ``matter''
within a fixed causal set, treated as a non-dynamical background or
``arena'' (in other words a discrete analog of quantum field theory in
curved spacetime).
As discussed in the introduction, most of our phenomenological insights
so far --- with the exception of those relating to $\Lambda$ --- stem
from the study of dynamics in this second sense.

I have already spoken about work on the propagation of classical
particles and of classical fields through a causet, and in the next
section I will summarize some results on the special case of the latter
where the field is a massless scalar.  However, first I would like to
dwell briefly on the dynamics of the causet itself.

In addition, let me call attention here to some results on the
propagation of {\it quantum} particles in a causet that relate to the
topic of the next section [5] and that are
described more fully in the contribution of Steven Johnston to this
volume [19].
In a certain sense, the Green function $G$ of the next section already
has the character of a ``path integral'', not for fields but for
particles.  That is, it can be expressed as a sum over paths, each path
contributing a single ``field amplitude''.  The only thing is that these
``paths'' verge on triviality: each is a chain of just two linked
elements of the causet.  That it is nonetheless appropriate to regard
such chains as discrete worldlines becomes clearer when one generalizes
to the massive case.  In the continuum, the massive Green function has
support in the interior the light cone, not just on its surface, and one
finds correspondingly that the ``discrete worldlines'' now comprise far
more than two elements [19].

Let us proceed now to causet dynamics proper.  Perhaps the first
question one should ask in this connection is how such a dynamics could
even be formulated, given that Hamiltonian evolution can get no foothold
when time as well as space is discrete, and when there is no background
manifold to support any sort of analog of a metric-field operator.  What
springs to mind, of course, is the path integral, but even this is
problematic if one limits its role to furnishing an ``evolution
operator'' between fixed boundaries.  
Rather, one must (or so it seems) replace the ``path integral qua
propagator'' by the path integral qua decoherence functional or
quantal measure.\footnote{$^\star$}
{See [20] for definitions of the decoherence
 functional and the quantal measure, these being essentially equivalent
 objects.}
When expressed in such measure-theoretic language, quantum theory
resembles not so much a deterministic theory like classical mechanics
as a classical probabilistic theory like that of of the random walk.
It appears in fact as a generalization of the theory of stochastic
processes that replaces the Kolmogorov sum-rule with a weaker sum-rule
that allows for pairwise interference between alternatives
[20].

In light of this resemblance, it appeared natural to seek a quantal
``law of motion'' for causal sets by first setting oneself the
preliminary task of finding a classically stochastic theory of causet
structure or development.  In fact what emerged were not ``laws of
structure'' as such, but rather a dynamics based on {\it growth} or
accretion, in other words a birth process.  The resulting Markov
processes are known as classical sequential growth (CSG) models, and
they follow almost uniquely from two or three general principles,
including in particular a principle of ``Bell causality''.  (Aside from
some exceptional solutions, all of these models belong to the so called
``generalized percolation'' family, and as such are closely related to
what are called ``random graph orders'' in the mathematical literature
[21].)

Although the CSG models were initially intended only as stepping stones
to a full quantum dynamics --- a dynamics of ``{\it quantal} sequential
growth'' --- one can ask to what extent any of them can produce causal
sets that resemble spacetimes (ie that are well approximated by a
Lorentzian manifold, preferably a spacetime that also satisfies the
Einstein equations!)  The answer seems to depend on how strong a
resemblance one is looking for.  On one hand there are good reasons to
believe that fully manifold-like causets will never be produced by any
of the CSG models (more precisely, will be produced with negligible
probability).  On the other hand, for certain choices of the parameters,
one obtains causets that resemble manifolds in important ways,
e.g. deSitter space [22].

Thanks to this resemblance, the CSG models can bring to life
possibilities inherent in the causal set hypothesis, that might be
realized more fully (i.e. more realistically) in some future theory of
{\it quantal} sequential growth.  One such example concerns a
``Boltzmann-Tolman universe'' scenario that could solve some of the
``large number riddles'' of cosmology [23].  It turns
out that for a large range of their parameters or ``coupling
constants'', the CSG models lead to a qualitatively similar overall
behavior of the growing causet $C$.  It begins life as a single element
(or as the empty set if you prefer to start there), expands rapidly to a
large size, stabilizes, and after some time (depending on the parameter
values chosen initially) collapses back to a single element (called a
{\it post} in the combinatorial literature).  There ensues a succession
of cycles of expansion and contraction, separated by posts, each of
which appears to its descendants as a new ``big bang''.  In each
successive cycle the initial expansion looks at first exactly the same,
beginning with a ``tree phase'' in which each new element is born from a
single ``parent'' chosen at random from the previously born elements of
the growing tree.\footnote{$^\dagger$}
{A causet is a {\it tree} iff each element has exactly one parent (or
 none in the case of the ``root'' element of the tree.)}
The subsequent phases of the expansion do depend on the specific values
of the parameters, but what is interesting is this, that the parameters
governing the evolution change from cycle to cycle.  As one cycle ends
and the next begins, the effective parameters are transformed by the
action of a certain ``cosmic renormalization group'' [24]
that drives them toward the values characteristic of the particular CSG
model known as {\it transitive percolation}, which contains a single
free parameter $p$ that, among other things, governs the size of maximum
expansion.  The value of this ``renormalized'' $p$ decreases from cycle
to cycle, and when it is much less than unity, $C$ expands to a very
large and homogeneous configuration.  Overall the picture seems
remarkably lifelike, given the limitations of classical growth.  In such
a scenario, the large numbers of cosmology, like the very great value of
the spatial radius of curvature in natural units, would owe their size
to the large age of the cosmos (measured in number of elements), rather
than to any special ``fine tuning''.\footnote{$^\flat$}
{The puzzle of why the radius of curvature is so much greater that the
 Planck length is often called the ``flatness problem''.}

The classical sequential growth models have also led to important
progress of a conceptual nature by showing how the so called ``problems
of time'' can be resolved in a way that applies equally well to the
classical and quantum cases [25].
Without any appeal to ``clocks'' or other surrogates for coordinate systems 
(like asymptotic regions where an S-matrix can be defined), 
one has identified ``observables'' which first
of all carry a clear physical meaning and second of all are complete in
the sense that all other covariant (label independent) quantities can
be formed from them in a well defined sense.  Because this completeness
theorem rests only lightly on the details of the CSG models, there is
every reason to expect that it will go through unchanged in the case of
quantal sequential growth (ie quantum gravity {\`a} la causet).

Of course this expectation can only be tested after we have a quantum
dynamics to work with.  How might one arrive at such a dynamics?  The
example of classical gravity suggests two very different but
complementary ways to proceed, one based on general principles the other
on imitation of the continuum theory.

The first approach was the one that brought the CSG models to light, the
analogy here being with how the Einstein equation was (or might have
been) deduced from the twin principles of general covariance and
locality.  As I stressed above, locality is not to be met with in a
causal set, and covariance by itself (in the sense of labeling
independence) did not seem to offer adequate guidance.  But it turned
out that a condition that came to be called ``Bell causality'' could
take over where locality left off.  The resulting CSG models are almost
uniquely selected by these two principles, together with a so-called
principle of ``internal temporality''.  (Full details of the derivation
can be found in [26].)  To derive in the same manner a
dynamics of ``quantal sequential growth'' has so far not been possible,
however, because one lacks a satisfactory generalization of the
condition of Bell causality to the quantum case.

The more general issue lurking here has nothing to do with quantum
gravity as such, and raises questions that are worthy of study in their
own right.  Can one formulate the requirement of relativistic causality
without invoking external agents like ``observers''?  Can one do so in
the language of histories?  And if so how?  The answers, if we had them
[27] [28],
would help to remove the main obstruction blocking the first path to a
theory of quantal sequential growth.
(The quest for a suitable criterion of relativistic causality lies in the
background of the work on quantum foundations reported on in Petros
Wallden's article in this volume [29].)

An approach based on imitation of the continuum theory might seem to be
more straghtforward than an approach based on first principles.  Along
this second way, one would try to discover a quantity in the causet that
would go over to the scalar curvature in the continuum limit.  Having
found such an expression, one would then try to incorporate it into a
discrete analog of the gravitational functional-integral.  For a long
time it was difficult to get started on this path because no discrete
counterpart of the curvature was known.  Now that we do have a candidate
expression (based on the Synge function as described earlier), the next
step would be to confirm that it really does have the Ricci scalar as
its continuum limit.\footnote{$^\star$}
{A more ``implicit'' way to arrive at a counterpart of the
 Einstein-Hilbert action has been suggested recently by Bombelli and
 Sverdlov [30].
 It is based on a kind of ``dynamical selection of a local frame'' that
 would allow one to deploy known estimators for certain components of
 the Ricci {\it tensor} in a causal set [31] [32].}

\section{3.~ direct transmission along causal links}
Earlier, I promised to return in this section to the model of
propagation of a scalar field by direct transmission along causal links
from source to sink.  In my actual talk at the DICE conference, I
devoted most of the time to this model, but since the details will
appear soon in another place [5], I will
limit myself here to a thumbnail sketch.

In the continuum, the field emitted by a specified source can be
computed in two ways, either by solving the field equations with
suitable boundary conditions or (for a free field) by folding the source
in with a suitable Green function.  Ordinarily one adopts the
``thermodynamic'' boundary condition of no incoming radiation and
correspondingly employs the retarded Green function.  What the model in
question does is to adapt the second of these two methods to the causet.

Now for a massless scalar field $\phi$ in $\Minkowski^4$, $G$ is
particularly simple, being given (up to normalization) just by the
delta-function on the future light cone, i.e. by $\delta(x^\mu x_\mu)$
in Cartesian coordinates (``Huygens' principle'').  In this sense,
$\phi$ propagates strictly between pairs of points joined by a null line
segment.  Now let $C$ be a causet well approximated by (or by a portion
of) $\Minkowski^4$.  The nearest discrete analog of a null segment is
perhaps a {\it link} of $C$, that is, a pair of elements $x\prec y$
related as parent and child, or in other words a pair such that the
interval delimited by $x$ and $y$ is empty.\footnote{$^\dagger$}
{A slightly more general analog would be a pair $x\prec y$ such that the
 interval between $x$ and $y$ is a chain.  In the (flat) continuum, this
 precisely characterizes a light ray. }
If, then, we define $G(x,y)=1$ or $0$ according as $x$ is or is not a
child of $y$, we obtain a natural analog of the continuum Green
function; and it is perhaps not too surprising that the continuum
limit of $G(x,y)$ coincides (up to normalization) with the 
retarded Green function in $\Minkowski^4$.   

To complete the model, we can represent the source as simply (though not
as realistically) as possible as a so called saturated chain or {\it
path} within $C$,\footnote{$^\flat$}
{A chain is saturated when no further elements can be interpolated
 between its endpoints.  In other words it is ``made exclusively of links''.}
and we assume that the value of $\phi(x)$ due to the source is, modulo a
prefactor of the source's charge, just the sum of $G(x,y)$ over all
elements $y$ of the source.  Clearly this is just the number of source
elements $y$ linked to $x$.  Finally we can imagine that the detector is
active in a specified region of spacetime (it has a certain spatial
volume and is turned on for a certain time) and just outputs the
integral of the field over the region.  Translated into causet terms,
this just means (because ``volume = number'') that the signal is the
{\it sum} of $\phi(x)$ over those $x\in{C}$ where the detector is
active.

Given this setup, it becomes relatively easy to compute the expectation
value of the detector's response to a given source, and also the leading
correction thereto in powers of the Planck length.  At leading order,
one recovers exactly the continuum expression,\footnote{$^\star$}
{In particular, (the AC part of) the field of an oscillating charge
 falls off, as it should, like $1/R$ at large distances, not as $1/R^2$ as
 one might have concluded naively from the fact that the DC part goes
 like $1/R$.  Interestingly, it is the Doppler shift in the number of
 contributing source elements that invalidates the naive reasoning.}
while the first
correction involves an integral over the boundary of the source region
and is completely negligible for any realistic detector and source.
Thus as expected, there is no change to the dispersion relations, and no
loss of coherence.  (Nor is there any scattering, but that conclusion
seems less certain to persist in the context of a more complete model
that attributes degrees of freedom to $\phi$ itself, and then deduces
the radiated field by solving the corresponding discrete field
equations.)

There remains the question of the signal fluctuations induced by the
discreteness of the underlying causet --- the fact that the signal is
actually a sum rather than an integral.  The fluctuations in $\phi$ are
less easy to compute than its expectation-value, and not every detail
has been pinned down yet.  However, it seems fairly clear that ---
unfortunately --- the fluctuations are also too tiny to show up in
observations of any realistic source using any realistic detector, even
if the source is at cosmological distances from us.

In summary, the model is one in which the field spreads via a process of
direct transmission along causal links.  It seems remarkable that this
law of propagation, which is probably the simplest one could have
imagined had one thought directly in terms of the causal set, turns out
to reproduce all the main features of the relationship between the field
and its source in the continuum.


\bigskip
\noindent
Research at Perimeter Institute for Theoretical Physics is supported in
part by the Government of Canada through NSERC and by the Province of
Ontario through MRI.
This research was partly supported 
by NSF grant PHY-0404646.

\ReferencesBegin                             

\ref [1] Rafael D.~Sorkin,
``Big extra dimensions make $\Lambda$ too small'',
 in the proceedings of the 
  Second International Workshop ``DICE 2004'',
   held September, 2004, Piombino, Italy,
    edited by Hans-Thomas Elze,
    \journaldata{Brazilian Journal of Physics}{35}{280-283}{2005},
    \eprint{gr-qc/0503057}

\ref [2] 
See for example 
A. P. Balachandran and M. Martone,
``Space-time from Symmetry: The Moyal Plane from the Poincar{\'e}-Hopf Algebra''
\eprint{arXiv:0902.3409 [hep-th]}

\ref [3] Luca Bombelli, Joe Henson and Rafael D. Sorkin,
``{Discreteness without symmetry breaking: a theorem}''
\eprint{gr-qc/0605006}

\ref [4] 
R.~Lieu and L.~W. Hillman, 
``The phase coherence of light from extragalactic sources: Direct evidence
 against first order quantum gravity fluctuations in time and space,'' 
\journaldata{Astrophys. J.}{585}{L77--L80} {2003}

\ref [5] 
Fay Dowker, Joe Henson and Rafael D.~Sorkin,
``Wave propagation on a causet I: direct transmission along causal links''
(in preparation)

\ref [6]
R.M Dudley,			  
``Lorentz-invariant Markov processes in relativistic phase space'',
\journaldata{Arkiv f{\"o}r Matematik}{6(14)}{241-268}{1965}
 %
\sepref
R.M Dudley,
``A note on Lorentz-invariant Markov processes'',\lbr
\journaldata{Arkiv f{\"o}r Matematik}{6(30)}{575-581}{1967}
 %
\sepref
G{\'e}za Schay, Jr., {\it The equations of diffusion in the Special Theory of Relativity}
(doctoral thesis, Princeton University, 1961)

\ref [7] 
Fay Dowker, Joe Henson and Rafael D.~Sorkin,
``Quantum Gravity Phenomenology, Lorentz Invariance and Discreteness'',
\journaldata {Modern Physics Letters~A} {19} {1829-1840} {2004}
\eprint{gr-qc/0311055},
\hpf{ http://www.physics.syr.edu/~sorkin/some.papers/ }

\ref [8]  
  Lydia Philpott, Fay Dowker, and Rafael D.~Sorkin,
  ``Energy-momentum diffusion from spacetime discreteness'',
  \eprint{arxiv:0810.5591 [gr-qc]}
  \lbr
  \hpf{ http://www.physics.syr.edu/~sorkin/some.papers/ }
\sepref
See also the article by Lydia Philpott in this volume. 

\ref [9]  
Nemanja Kaloper and David Mattingly,
``Low energy bounds on Poincar{\'e} violation in causal set theory'',
 \journaldata{Phys. Rev. D}{74}{106001}{2006},
 \eprint{arXiv:astro-ph/0607485}

\ref [10] 
Rafael D. Sorkin,
``Does Locality Fail at Intermediate Length-Scales?''
 in  {\it Approaches to Quantum Gravity -- Towards a new understanding of space and time}, 
 edited by Daniele Oriti 
 (Cambridge University Press 2009)
 (ISBN: 978-0-521-86045-1), pages 26-43,
 \eprint{gr-qc/0703099}, \ 
 \hpf{ http://www.physics.syr.edu/~sorkin/some.papers/ }
\sepref
 Joe Henson, ``The causal set approach to quantum gravity''
 \eprint{gr-qc/0601121}

\ref [11] 
 Rafael D.~Sorkin, 
 ``A Modified Sum-Over-Histories for Gravity'',
   reported in
  {\it 
   Highlights in gravitation and cosmology: 
   Proceedings of the International Conference on Gravitation and Cosmology, 
   Goa, India, 14-19 December, 1987},
   edited by 
   B.~R. Iyer, Ajit Kembhavi, Jayant~V. Narlikar, and C.~V. Vishveshwara,
   see pages 184-186 in the article by 
   D.~Brill and L.~Smolin: 
   ``Workshop on quantum gravity and new directions'', pp 183-191 
   (Cambridge University Press, Cambridge, 1988)
\sepref
``On the Role of Time in the Sum-over-histories Framework for Gravity'',
    paper presented to the conference on 
    The History of Modern Gauge Theories, 
    held Logan, Utah, July 1987, 
    published in  
    \journaldata {Int. J. Theor. Phys.}{33}{523-534}{1994}

\ref [12]  
Maqbool Ahmed, Scott Dodelson, Patrick Greene and Rafael D.~Sorkin,
``Everpresent $\Lambda$'',
\journaldata {Phys. Rev.~D} {69} {103523} {2004},
\eprint{astro-ph/0209274}, 
\lbr
\hpf{ http://www.physics.syr.edu/~sorkin/some.papers/ }
\sepref
Maqbool Ahmed, 			
``First indications of causal set cosmology''
Doctoral dissertation
(Syracuse University, 2006) 
\sepref
Y.~Jack Ng and H.~van Dam, ``A small but nonzero cosmological constant'',
\journaldata{Int. J. Mod. Phys D.}{10}{49}{2001}
\eprint{hep-th/9911102}
\sepref
Rafael D.~Sorkin,
`` Is the cosmological ``constant'' a nonlocal quantum residue of
   discreteness of the causal set type?'',
in the proceedings of the PASCOS-07 Conference, held July, 2007, London, England,
American Institute of Physics Conference Proceedings 957, 142-153, 
ISBN: 978-0-7354-0471-7  (2007),\lbr
\arxiv{0710.1675 [gr-qc]},\lbr
\hpf{ http://www.physics.syr.edu/~sorkin/some.papers/ }

\ref [13]
Luca Bombelli, Joohan Lee, David Meyer and Rafael D.~Sorkin,
``Spacetime as a Causal Set'',
  \journaldata {Phys. Rev. Lett.}{59}{521-524}{1987}
\sepref
Luca Bombelli, {\it Space-time as a Causal Set},  
  Ph.D. thesis, Syracuse University (1987)
\sepref
Rafael D.~Sorkin,
``Causal Sets: Discrete Gravity (Notes for the Valdivia Summer
School)'',
in {\it Lectures on Quantum Gravity}
(Series of the Centro De Estudios Cient{\'\i}ficos),
proceedings of the Valdivia Summer School,
held January 2002 in Valdivia, Chile,
edited by Andr{\'e}s Gomberoff and Don Marolf
(Plenum, 2005)
\eprint{gr-qc/0309009}
\sepref
Fay Dowker, ``Causal sets and the deep structure of Spacetime'',
 in
 {\it 100 Years of Relativity - Space-time Structure: Einstein and
 Beyond}", p445-464
 ed Abhay Ashtekar
 (World Scientific, 2005)
 \eprint{gr-qc/0508109}
\sepref
Rafael D. Sorkin, ``Geometry from order: causal sets'',
  online article in the {\it Spotlights on relativity} series,
  maintained by
  the Max-Planck-Institut f{\"u}r Gravitationsphysik
  (Albert-Einstein-Institut), Potsdam,\lbr
 http://www.einstein-online.info/en/spotlights/causal\_sets/index.html
\sepref
Joe Henson, ``The causal set approach to quantum gravity''
 \eprint{gr-qc/0601121}

\ref [14]
Djamel Dou,			
 ``Causal Sets, a Possible Interpretation for the Black Hole
 Entropy, and Related Topics'', 
 Ph.~D. thesis (SISSA, Trieste, 1999)
 \eprint{gr-qc/0106024}
\sepref
Djamel Dou and Rafael D.~Sorkin, ``Black Hole Entropy as Causal Links'',
\journaldata {Foundations of Physics}{}{}{to appear}
\eprint{gr-qc/0302009}

\ref [15] D.A.~Meyer, {\it The Dimension of Causal Sets}. Ph.D. thesis, M.I.T. (1988)

\ref [16]
Eitan Bachmat, private communication.

\ref [17]
See the contributions by Sumati Surya and David Rideout in this volume
for these developments and for further references.

\ref [18]
Seth Major, David Rideout, and Sumati Surya
``On Recovering Continuum Topology from a Causal Set''
\eprint{gr-qc/0604124}

\ref [19] 
Contribution by Steven Johnston in this volume.

\ref [20] 
Hartle, J.B., ``The Quantum Mechanics of Cosmology'', 
 in {\it Quantum Cosmology and Baby Universes: Proceedings of the 1989
 Jerusalem Winter School for Theoretical Physics}, eds. S.~Coleman et al. 
 (World Scientific, Singapore, 1991)
\sepref
Rafael D.~Sorkin,
``Quantum Mechanics as Quantum Measure Theory'',
   \journaldata{Mod. Phys. Lett.~A}{9 {\rm (No.~33)}}{3119-3127}{1994}
   \eprint{gr-qc/9401003}

\ref [21]
Graham Brightwell, ``Models of Random Partial Orders'',   
 in {\it Surveys in Combinatorics, 1993},
 London Math. Soc. Lecture Notes Series {\bf 187}:~53-83,
 ed. Keith Walker
 (Cambridge Univ. Press 1993)

\ref [22]  
Maqbool Ahmed and David P. Rideout (unpublished).

\ref [23] 
Rafael D.~Sorkin,
``Indications of causal set cosmology'',
 \journaldata {Int. J. Theor. Ph.} {39 {\rm(7)}} {1731-1736} {2000}
 (an issue devoted to the proceedings of the Peyresq IV conference,
  held June-July 1999, Peyresq France),
 \eprint{gr-qc/0003043},
 \lbr
 \hpf{ http://www.physics.syr.edu/~sorkin/some.papers/ }

\ref [24] 
Xavier Martin, Denjoe O'Connor, David Rideout and Rafael D.~Sorkin,
``On the `renormalization' transformations induced by
  cycles of expansion and contraction in causal set cosmology'',
 \journaldata{Phys. Rev.~D}{63}{084026}{2001},
 \eprint {gr-qc/0009063},
 \lbr
 \hpf{ http://www.physics.syr.edu/~sorkin/some.papers/ }
\sepref
Avner Ash and Patrick McDonald,
``Moment problems and the causal set approach to quantum gravity'',
\journaldata{J. Math. Phys.}{44}{1666-1678}{2003}
\eprint{gr-qc/0209020}

\ref [25] 
Graham Brightwell, {H. Fay Dowker}, {Raquel S. Garc{\'\i}a}, {Joe Henson} 
 and {Rafael D.~Sor\-kin},
``General Covariance and the `Problem of Time' in a Discrete Cosmology'',
 in K.G.~Bowden, Ed., 	
 {\it Correlations}, 
 Proceedings of the ANPA 23 conference,
 held August 16-21, 2001, Cambridge, England 
 (Alternative Natural Philosophy Association, London, 2002), pp 1-17
\eprint{gr-qc/0202097}
\sepref
Graham Brightwell, Fay Dowker, Raquel S.~Garc{\'\i}a, Joe Henson and 
     Rafael D.~Sorkin,
``{$\,$}`Observables' in Causal Set Cosmology'',
\journaldata {Phys. Rev.~D} {67} {084031} {2003}\lbr
\eprint{gr-qc/0210061}

\ref [26] David P.~Rideout, {\it Dynamics of Causal Sets}, Ph.D. thesis
 (Syracuse University 2001)
\sepref
David P.~Rideout and Rafael D.~Sorkin,
``A Classical Sequential Growth Dynamics for Causal Sets'',
 \journaldata{Phys. Rev.~D}{61}{024002}{2000}
 \eprint{gr-qc/9904062}

\ref [27] 
Joe Henson, ``Comparing causality principles'',
\journaldata{Stud. Hist. Philos. Mod. Phys.}{36}{519-543}{2005}
\eprint{quant-ph/0410051}

\ref [28] 
David Craig, Fay Dowker, Joe Henson, Seth Major, David Rideout and Rafael D.~Sorkin,
``A Bell Inequality Analog in Quantum Measure Theory'',
\journaldata{J. Phys. A: Math. Theor.}{40}{501-523}{2007},
\eprint{quant-ph/0605008},
\lbr
\hpf{ http://www.physics.syr.edu/~sorkin/some.papers/ }

\ref [29] 
Contribution by Petros Wallden in this volume.

\ref [30]
See the contribution by Luca Bombelli in this volume, 
also \eprint{arXiv:0905.1506}.



\ref [31] J.~Myrheim, ``Statistical geometry'', CERN preprint TH-2538 (1978)
(available from SPIRES)

\ref [32] 
G. W. Gibbons and S. N. Solodukhin, 
``The Geometry of Small Causal Diamonds''
http://arxiv.org/abs/hep-th/0703098

\end               


(prog1    'now-outlining
  (Outline 
     "\f......"
      "
      "
      "
   ;; "\\\\message"
   "\\\\Abstrac"
   "\\\\section"
   "\\\\subsectio"
   "\\\\appendi"
   "\\\\Referen"
   "\\\\ref....[^|]"
  ;"\\\\ref....."
   "\\\\end